\documentclass[12pt]{iopart}
\usepackage{bm}
\usepackage{graphicx}
\usepackage{color}

\begin{document}

\title[] {Structure, magnetism and electronic properties in 3$d$-5$d$ based double perovskite (Sr$_{1-x}$Y$_x$)$_2$FeIrO$_6$}

\author{K C Kharkwal$^1$ and A K Pramanik$^1$}

\address{$^1$ School of Physical Sciences, Jawaharlal Nehru University, New Delhi - 110067, India}

\eads{\mailto{akpramanik@mail.jnu.ac.in}}

\begin{abstract}
The 3$d$-5$d$ based double perovskites are of current interest as they provide model system to study the interplay between electronic correlation ($U$) and spin-orbit coupling (SOC). Here we report detailed structural, magnetic and transport properties of doped double perovskite material (Sr$_{1-x}$Y$_x$)$_2$FeIrO$_6$ with $x$ $\leq$ 0.2. With substitution of Y, system retains its original crystal structure but structural parameters modify with $x$ in nonmonotonic fashion. The magnetization data for Sr$_2$FeIrO$_6$ show antiferromagnetic type magnetic transition around 45 K, however, a close inspection in data indicates a weak magnetic phase transition around 120 K. No change of structural symmetry has been observed down to low temperature, although the lattice parameters show sudden changes around the magnetic transitions. Sr$_2$FeIrO$_6$ shows an insulating behavior over the whole temperature range which yet does not change with Y substitution. Nature of charge conduction is found to follow thermally activated Mott's variable range hopping and power law behavior for parent and doped samples, respectively. Interestingly, evolution of structural, magnetic and transport behavior in (Sr$_{1-x}$Y$_x$)$_2$FeIrO$_6$ is observed to reverse with $x$ $>$ 0.1 which is believed to arise due to change in transition metal ionic state. 
\end{abstract}


\section {Introduction}

Recently, a lot of interest has been given to 3$d$-5$d$ based double perovskite (DP) systems A$_2$BB$'$O$_6$, where A is the alkaline- or rare-earth element and B/B$'$ are the 3$d$/5$d$ transition metal (TM) elements. The interest in these materials mainly comes as 3$d$ TM has large electronic correlation ($U$) which is substantially reduced in 5$d$ element whereas the spin-orbit coupling (SOC) effect becomes sufficiently stronger with increasing $d$ character from 3$d$ to 5$d$ elements. Therefore, the interplay between $U$ and SOC, which remains largely an unexplored area and can induce many interesting and novel physical properties where some of those are of technological use, can be studied in controlled fashion in these materials.\cite{pardo,gupta,pesin,balents,vasala,chen,krock,philipp,taylor,kartik,zhu1} Till date, main research in these compounds has mainly focused on studying the effect with variation of B and B$'$ elements as well as partially replacing the A-site element with either divalent or trivalent cation. For TM elements, till so far cations such as, B = Fe, Co, Ni, Mg, Zn, Cu, Y, etc. and B$'$ = Re, Os and Ir have been used. Most of the 3$d$-5$d$ based DP systems show insulating electronic transport behavior over the temperature. From point of magnetic ordering, these materials mostly show long-range type antiferromagnetic (AFM) spin structure at low temperature.\cite{laghuna,kayser,veiga,teresa,narayanan,poddar, kolchi,kato,rene,rolf,kayser1} In addition, the change of spin structure with temperature in some of these systems indeed is an interesting observation.\cite{avijit,williams,yan}
  
The Ir based DP materials have been the topic of intense research in recent times because of its unusual physical properties which arise from its extended 5$d$ orbital. Under the crystal field effect (CFE), the 5$d$ orbitals of Ir splits into higher energy $e_g$ and lower energy $t_{2g}$ states. It is believed that strong SOC again splits $t_{2g}$ state into $J_{eff}$ = 3/2 quartet and 1/2 doublet states.\cite{kim}  This plays an interesting role in deciding the magnetic and transport properties for Ir containing materials. For instance, most commonly Ir adopts oxidation states of Ir$^{4+}$ and Ir$^{5+}$ which has electronic configuration 5$d^5$ and 5$d^4$, respectively. With strong CEF, five electrons of Ir$^{4+}$ adopts a low spin state where four of those fill $J_{eff}$ = 3/2 state while remaining one partially occupy $J_{eff}$ = 1/2 state. This gives single-spin (magnetic) $J_{eff}$ = 1/2 picture giving interesting physics in materials of interests such as, layered perovskites Sr$_{2}$IrO${_4}$,\cite{kim,kim1,bhatti} double perovskites La$_2$ZnIrO$_6$, La$_2$MgIrO$_6$, etc.\cite{cao,zhu} On the other hand, four $d$-electrons of Ir$^{5+}$ with completely fill $J_{eff}$ = 3/2 state, henceforth, they are considered to be nonmagnetic ($J_{eff}$ = 0). Interestingly, nonmagnetic ground state, at least till sufficiently low temperature, has recently been shown for Sr$_2$YIrO$_6$ and Ba$_2$YIrO$_6$ DP systems where Y and Ir takes up ionic state of Y$^{+3}$ and Ir$^{+5}$ which are arguably nonmagnetic.\cite{dey,cao1} However, these studies further show signature of magnetic transition and/or magnetic moment at very low temperature in these DP materials below 1.3 and 0.4 K, respectively which arises from Ir$^{+5}$ ions where the origin of magnetism has mainly been attributed to strong noncubic crystal fields.  

Sr$_2$FeIrO$_6$ is an interesting member of 3$d$-5$d$ based DP family, where TMs are believed to be in Fe$^{3+}$ and Ir$^{5+}$ charge state with 3$d^5$ and 5$d^4$ electronic configuration, respectively. This implies magnetism in these materials is realized only through Fe-O-Fe network. While this material is an insulator throughout the temperature range, previous magnetization study shows a broad cusp or maximum around 120 K which signifies a transition to long-range type AFM state at low temperature.\cite{battle,bufaical,qasim} Crystal structure of Sr$_2$FeIrO$_6$ is though debated but the previous studies mostly show a distorted monoclinic or triclinic structure at room temperature.\cite{battle,bufaical,qasim} Nonetheless, understanding the evolution of crystal structure with temperature, in particular across the magnetic phase transition would be interesting for this material.   

In this present work, we have studied detailed structural, magnetic and transport properties of doped (Sr$_{1-x}$Y$_x$)$_2$FeIrO$_6$ ($x$ $\leq$ 0.2) DP system. Yttrium (Y$^{3+}$) is non-magnetic, therefore, it excludes the possibility of $f$-$d$ exchange interaction and does not add further complication in magnetic interaction. Yttrium substitution at Sr-site mainly has two effects; one is related to ionic size mismatch between Y$^{3+}$ (0.96 \AA) and Sr$^{2+}$ (1.44 \AA) which would lead to structural modification and another is change of charge state of either Fe or Ir. We find undoped Sr$_2$FeIrO$_6$ crystallizes in triclinic structure, and with Y substitution even though structural parameters change in nonmonotonic fashion but the original structural symmetry is retained. Magnetization data show prominent signature of phase transition around 45 K, however, a similar though weak feature has also been observed around 120 K. With lowering of temperature no structural phase transition has been observed around magnetic transition, but the lattice parameters show changes around magnetic transition. Sr$_2$FeIrO$_6$ is insulating over whole temperature range where the resistivity follows variable range hopping behavior at two distinct temperature range. Substitution of Y on Sr-site in (Sr$_{1-x}$Y$_x$)$_2$FeIrO$_6$ has interesting effect where structural, magnetic and transport properties evolve with $x$, however, the nature of change reverses for $x$ $>$ 0.1.

\section {Experimental Details}

Polycrystalline samples of series (Sr$_{1-x}$Y$_x$)$_2$FeIrO$_6$ with $x$ = 0.0, 0.05, 0.1, 0.15 and 0.2 have been prepared using solid state reaction method. The high purity ingredient powders of SrCO$_3$, Fe$_2$O$_3$, IrO$_2$ and Y$_2$O$_3$ are taken in stoichiometric ratio and ground well. The Y$_2$O$_3$ has been given pre-heat treatment at 800${^\circ}$C for 8 hours to remove added moisture. The fine powders are calcined in air at 900${^\circ}$C for 24 hours twice with an intermediate grinding. Calcined powders are then palletized and sintered at 1000${^\circ}$C, 1050${^\circ}$C and 1100${^\circ}$C for 24 hours intermediate grindings. The phase purity of these materials has been checked using powder x-ray diffraction (XRD) with a Rigaku MiniFlex diffractomer and CuK$_{\alpha}$ radiation at room temperature. The XRD data are refined with Rietveld program for structural analysis. The temperature dependent XRD measurements are done using a PANalytical powder diffractometer in the temperature range between 20 to 300 K. A helium based closed cycle refrigerator (CCR) is used to achieve low temperatures. Proper care has been taken for temperature stabilization by giving sufficient wait time before collecting data. Data are collected in 2$\theta$ range of 10 - 90${^\circ}$ at a step of 0.033${^\circ}$ and with a scan rate of 2${^\circ}$/min. DC magnetization data have been collected using a vibrating sample measurement (PPMS, Quantum Design) and electrical transport properties have been measured using a home-built insert attached with CCR.

\begin{figure}
\centering
		\includegraphics[width=8cm]{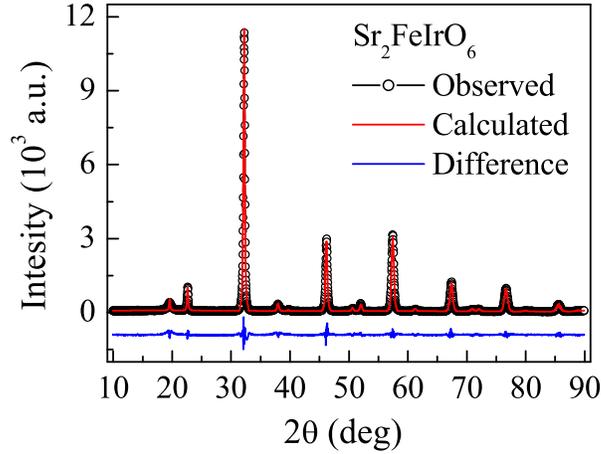}
\caption{(color online) XRD pattern of Sr$_2$FeIrO$_6$ collected at room temperature along with Rietveld refinement.}
	\label{fig:Fig1}
\end{figure}

\section{Results and Discussions}
\subsection{Structural analysis of (Sr$_{1-x}$Y$_x$)$_2$FeIrO$_6$ series}

In general, transition metal octahedra i.e., BO$_6$ and B$'$O$_6$ in DP materials are arranged alternatively in two interpenetrating sublattices.\cite{serrate} The structural symmetry, however, in these materials is intimately linked with the ionic sizes of constituent cations (A, B and B$'$) or rather with tolerance factor ($t$). Previous studies such as, M$\ddot{o}$ssbauer spectroscopy (MS),\cite{battle,bufaical} x-ray absorption spectroscopy (XAS)\cite{kayser} and x-ray magnetic circular dichroism (XMCD)\cite{laghuna} have conclusively indicated transition metal cations in Sr$_2$FeIrO$_6$ are in form Fe$^{3+}$ and Ir$^{5+}$, the fact which is also in agreement with our magnetization data. Using the ionic radii of corresponding ions we calculate $t$ = 1.01 which indicates this material would adopt a cubic structure. While considering the structural parameters as obtained from analysis of our XRD data, we calculate $t$ = 0.95 which suggests a distorted lattice structure i.e., monoclinic or triclinic structure for Sr$_2$FeIrO$_6$.\cite{serrate} Previous studies indeed show a distorted structure for Sr$_2$FeIrO$_6$, but there exists some disagreement over the structural phase symmetry. For instance, Battle \textit{et al.} \cite{battle} has shown monoclinic-\textit{P2$_1$/n} structure from XRD data while their high-resolution neutron diffraction data could be better fitted with triclinic-\textit{I$\bar{1}$} structure. Similarly, other groups have shown monoclinic-\textit{P2$_1$/n} structure using laboratory based XRD data\cite{bufaical} and monoclinic-\textit{I2/m} symmetry using synchrotron XRD and powder neutron diffraction data.\cite{qasim}

\begin{figure}
\centering
		\includegraphics[width=8cm]{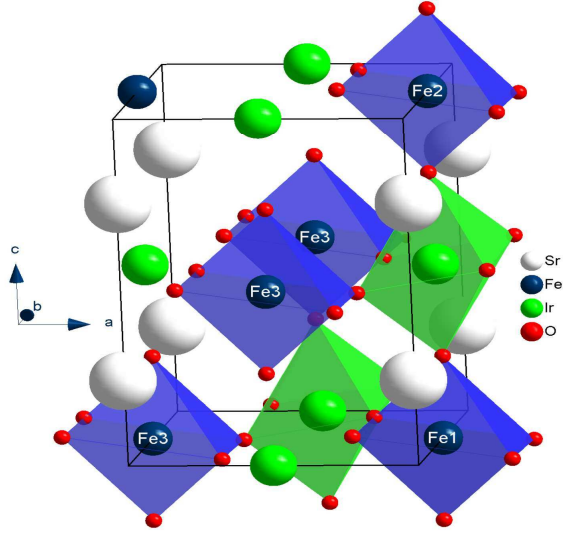}
\caption{(color online) Crystal structure of Sr$_2$FeIrO$_6$ unit cell, Blue and Green polyhedra represent FeO$_6$ and IrO$_6$ octahedra, respectively.}  
	\label{fig:Fig2}
\end{figure}

In our study, we have initially tried to refine our XRD data for Sr$_2$FeIrO$_6$ with above mentioned monoclinic and triclinic models. We found our XRD data could be better fitted with triclinic structure with \textit{I$\bar{1}$} symmetry as evidenced from lowest $\chi^2$ values. For example, we obtain $\chi^2$ values for triclinic-\textit{I$\bar{1}$}, monoclinic-\textit{P2$_1$/n} and monoclinic-\textit{I2/m} structures are 4.78, 5.28, 5.17, respectively. Fig. 1 shows XRD pattern along with  Rietveld analysis with triclinic-\textit{I$\bar{1}$} structure for parent Sr$_2$FeIrO$_6$ material. The Rietveld refinement of XRD data show the material is in single phase without any chemical impurity and crystallizes in triclinic-\textit{I$\bar{1}$} symmetry with lattice constants $a$ = 5.5514, $b$ = 5.5784, $c$ = 7.8435 \AA, and angels $\alpha$ = 90.092, $\beta$ = 89.866 and $\gamma$ = 89.960 $\deg$. Fig. 2 depicts unit cell atomic arrangement of triclinic-\textit{I$\bar{1}$} structure for Sr$_2$FeIrO$_6$ showing FeO$_6$ and IrO$_6$ octahedra are corner-shared and alternatively arranged in unit cell. It is clear in figure that there are two possible path of magnetic interactions: one is linear path between Fe1 and Fe2 through Ir along the $c$ axis (Fe1-O-Ir-O-Fe2) and another form a nonlinear path between Fe1 and Fe3 (Fe1-O-O-Fe3). The separation between Fe1-Fe2 and Fe1-Fe3 is $\sim$ 7.84 and 5.55 \AA, respectively. Therefore, overall magnetic and transport properties would be governed by the path of interaction which is chosen by the system.

\begin{figure}
\centering
		\includegraphics[width=8cm]{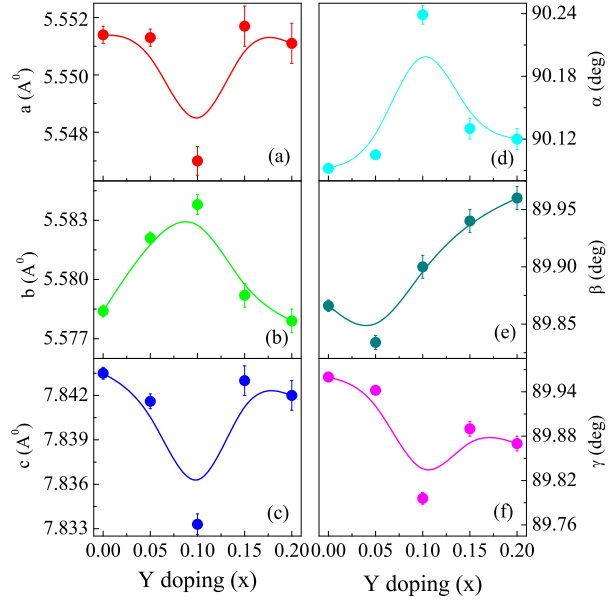}
\caption{(color online) Lattice parameters of triclinic unit cell (a) $a$ (b) $b$ (c) $c$ (d) $\alpha$ (e) $\beta$ and (f) $\gamma$ have been shown as a function of Y concentration $x$ for (Sr$_{1-x}$Y$_x$)$_2$FeIrO$_6$. Lines are guide to eyes.}
	\label{fig:Fig3}
\end{figure}

Rietveld analysis of XRD data for all the samples in present series show with substitution of yttrium, structural symmetry does not change and the system retains its original triclinic-\textit{I$\bar{1}$} symmetry. In Fig. 3, we have shown  unit cell parameters (lattice constants $a$, $b$ and $c$ and angels $\alpha$, $\beta$ and $\gamma$) for Sr$_2$FeIrO$_6$ and its evolution with $x$ in (Sr$_{1-x}$Y$_x$)$_2$FeIrO$_6$. It is evident in figure that evolution of lattice parameters is not monotonic with $x$ where the parameters initially decreases or increases but the nature of change reverses around $x$ = 0.1. Note, that similar nonmonotonic variation of structural parameters has also been observed with La$^{3+}$ substitution in present DP material i.e., in (Ca,Sr)$_{2-x}$La$_x$FeIrO$_6$.\cite{bufaical} In parent Sr$_2$FeIrO$_6$, cations are believed to be in form of Sr$^{2+}$, Fe$^{3+}$ and Ir$^{5+}$ with ionic radii 1.44, 0.645 and 0.57 \AA, respectively. The subsequent changes in ionic state of Fe/Ir with substitution of Y$^{3+}$ for Sr$^{2+}$ will have influence on the evolution of lattice parameters in (Sr$_{1-x}$Y$_x$)$_2$FeIrO$_6$. For instance, Y$^{3+}$ substitution will either convert Fe$^{3+}$ to Fe$^{2+}$ (0.78 \AA) or Ir$^{5+}$ to Ir$^{4+}$ (0.625 \AA). The large mismatch between ionic radii of Sr$^{2+}$ (1.44 \AA) and Y$^{3+}$ (0.96 \AA) will introduce further distortion in oxygen octahedra to accommodate the misfit between ionic radii of A and B/B$'$ atoms and will also reduce the unit cell volume. The subsequent conversion of both Fe$^{3+}$ to Fe$^{2+}$ or Ir$^{5+}$ to Ir$^{4+}$ will then increase the average ionic radii of B/B$'$ atoms and as a result distortion in octahedra will be reduced. Indeed, we find unit cell volume ($V$) initially decreases and then start to increase for $x$ $>$ 0.1 (not shown), following similar pattern of $a$ and $c$ (Figs. 3a and 3c). Analysis of our magnetization data imply that with initial substitution of Y$^{3+}$, mostly Fe$^{3+}$ converts to Fe$^{2+}$ up to $x$ $\sim$ 0.1, and after that Ir$^{5+}$ converts to Ir$^{4+}$ (discussed later). We believe that this change in ionic state with $x$ is responsible for change in evolution pattern of structural parameters around $x$ = 0.1 as seen in Fig. 3.
 
\begin{figure}
\centering
		\includegraphics[width=8cm]{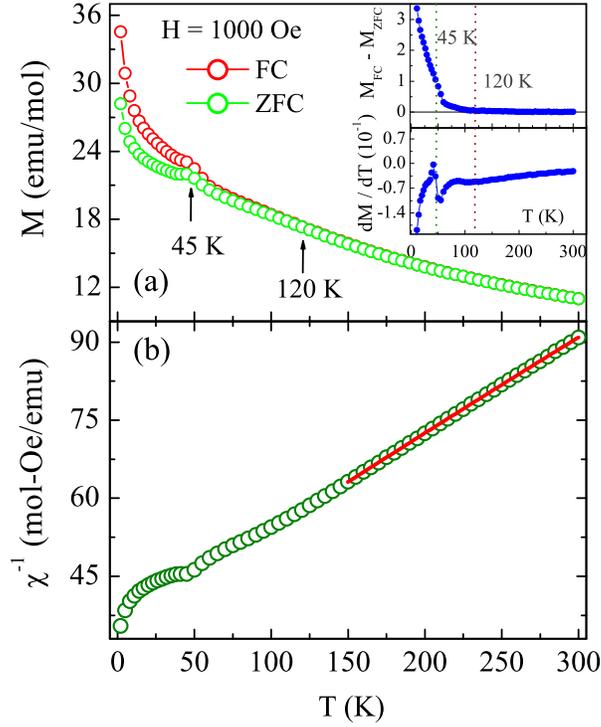}
\caption{(color online) (a) DC magnetization data measured in applied field of 1 kOe under ZFC and FC protocol are shown as a function of temperature for Sr$_2$FeIrO$_6$. Upper inset shows temperature dependence of difference between ZFC and FC magnetization data as shown in main panel. Lower inset shows temperature derivative of ZFC magnetization data. Two vertical dotted lines mark for possible two magnetic transition. (b) Temperature dependent inverse magnetic susceptibility ($\chi^{-1}$ = $(M/H)^{-1}$) deduced from ZFC magnetization data has been shown for Sr$_2$FeIrO$_6$. Solid line is due to fitting with Eq. 1 (discussed in text).}
	\label{fig:Fig4}
\end{figure}

\subsection{Magnetization study for parent Sr$_2$FeIrO$_6$}

Fig. 4a shows temperature ($T$) dependence of dc magnetization ($M$) data for Sr$_2$FeIrO$_6$ measured in 1000 Oe magnetic field following zero field cooled (ZFC) and field cooled (FC) protocol. On cooling, both $M_{ZFC}$ and $M_{FC}$ show a continuous increase over the temperature. However, the figure shows a weak kink at temperature $\sim$ 45 K in magnetization data, and below this temperature a clear bifurcation is observed between ZFC and FC magnetization. Though the magnetic property is not very extensively studied for Sr$_2$FeIrO$_6$, the nature of our $M(T)$ data is apparently different from reported studies. While some groups show a pronounced hump around 120 K and a bifurcation between $M_{ZFC}$ and $M_{FC}$ below $\sim$ 45 K,\cite{battle,bufaical} others show a wide bifurcation between $M_{ZFC}$ and $M_{FC}$ below $\sim$ 120 K.\cite{kayser,qasim} Nonetheless, reported studies show disagreement over the nature of magnetic behavior. To have a closer look, we have plotted difference between both magnetization data ($M_{FC}$ - $M_{ZFC}$) and derivative of ZFC magnetization (d$M$/d$T$) in upper and lower inset of Fig. 4a, respectively. It is clear in inset of Fig. 4a that onset of bifurcation between $M_{ZFC}$ and $M_{FC}$ originally starts at $\sim$ 120 K which becomes very prominent once system is cooled below 45 K. Similarly, d$M$/d$T$ shows a prominent inflection in $M(T)$ around 45 K but a change in curvature is also observed around 120 K. This indicates present Sr$_2$FeIrO$_6$ has long-range type magnetic transition at $\sim$ 45 K, yet onset of this magnetic transition could be at $\sim$ 120 K. Alternatively, the system may have original magnetic transition at $\sim$ 120 K and transition at $\sim$ 45 K is marked by change of spin structure. Note, that unlike metamagnetic transition such as, glass or superparamagnetic transition our $M(T)$ data do not show any cusp at low temperature around 45 K. It is worth to mention here that few of 3$d$-5$d$ based DP materials i.e., Sr$_2$FeOsO$_6$ and Sr$_2$CoOsO$_6$ have shown two magnetic phase transition temperatures related with change of spin structures well within AFM state.\cite{avijit, williams,yan,avijit1,sudipta,morrow} Needless to mention, understanding the similar types of magnetic transitions in Sr$_2$FeIrO$_6$ requires detailed investigation using local probes, yet the magnetization data in Fig. 4a appears to be interesting.

Fig. 4b shows temperature dependent inverse susceptibility ($\chi$ = $M/H$) deduced from ZFC magnetization data presented in Fig. 4a. In high temperature regime, $\chi^{-1}(T)$ shows linear behavior, however, a weak and pronounced kink in $\chi^{-1}(T)$ is observed around 120 and 45 K, respectively. In high temperature, data can be fitted with modified Curie-Weiss behavior,

\begin{eqnarray}
	\chi = \chi_0 + \frac{C}{T-\theta_P}
\end{eqnarray}

where $C$ and $\theta_{P}$ are the Curie constant and Curie temperature, respectively. From straight line fitting we have obtained $\chi_0$ = 6.9 $\times$ $10^{-4}$ emu/mol-Oe, $C$ = 4.83 and $\theta_{P}$ = -169 K. The value of $\theta_{P}$ is higher than the corresponding temperature scale (120 or 45 K), and also its negative sign is suggestive of non-ferromagnetic type spin ordering. While the low temperature magnetic state in Sr$_2$FeIrO$_6$ is believed to be of AFM type but this material possesses high value of frustration as indicated by large value of frustration parameter, $f$ = $|\theta_P|$/T$_N$ ($\sim$ 3.4). Using the Curie constant, we have calculated the effective paramagnetic moment ($\mu_{eff}$) which comes out to be 6.19 $\mu_B$/f.u. Using this obtained value of $\mu_{eff}$, we have tried to understand the possible ionic states of Fe and Ir. Out of different combination of Fe and Ir ionic states, we observe that pair of Fe$^{3+}$ and Ir$^{5+}$ gives expected $\mu_{eff}$ (=$\sqrt{S(S+1)}\mu_{B}$, where $S$ is total spin of atom) which is quite close to our experimentally observed $\mu_{eff}$ value (6.19 $\mu_B$/f.u). Fe$^{3+}$ with 3$d^5$ electronic state has $S$ = 5/2 in high-spin configuration which corresponds to $\mu_{eff}$ = 5.92 $\mu_{B}$/f.u. On the other hand, 5$d^4$ electronic state in Ir$^{5+}$ implies four $d$-electrons will fully occupy $J_{eff}$ = 3/2 quartet state ($J_{eff}$ = 0) which means $\mu_{eff}$ = 0. From these, we calculate effective $\mu_{eff}$ = $\left(\sqrt{(\mu_{eff}^{Fe})^2 + (\mu_{eff}^{Ir})^2}\right)$ = 5.92 $\mu_B$/f.u which is close to obtained value 6.19 $\mu_B$/f.u. Here, we mention that previous spectroscopy measurements (MS, XAS and XMCD) have indicated the Fe$^{3+}$ and Ir$^{5+}$ charge state in Sr$_2$FeIrO$_6$.\cite{battle,bufaical,kayser,laghuna} In this respect, our analysis of cation state is in agreement with microscopic investigations.

\begin{figure}
\centering
		\includegraphics[width=8cm]{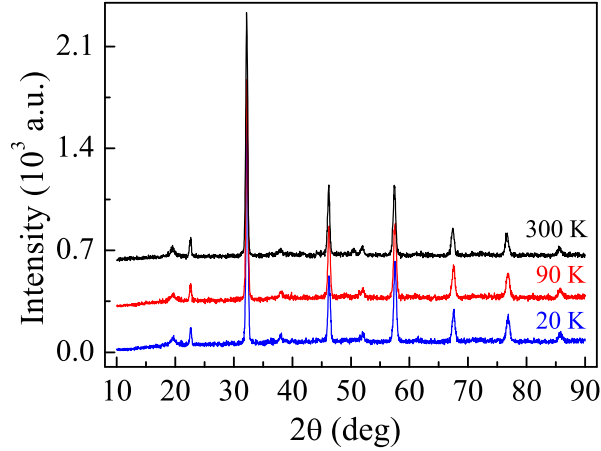}
\caption{(color online) XRD pattern have been shown for Sr$_2$FeIrO$_6$ at three selective temperatures 20, 90 and 300 K. Pattern at 90 and 300 K has been shifted vertically for clarity.}
	\label{fig:Fig5}
\end{figure}  
 
\subsection{Temperature dependent structural analysis for parent Sr$_2$FeIrO$_6$}

As the magnetic measurements for Sr$_2$FeIrO$_6$ show a ordered magnetic phase at low temperature, now to understand whether this magnetic phase transition is related to structural modification we have done temperature dependent structural analysis using XRD. Fig. 5 shows representative XRD pattern for Sr$_2$FeIrO$_6$ at 300, 90 and 20 K where the temperatures represent high temperature paramagnetic state, possible magnetic state in-between 120 and 45 K and low temperature ordered magnetic state. We find no significant changes in XRD pattern over the temperature in terms of peak position and/or peak splitting which primarily implies magnetic phase transition is not accompanied with structural phase transition. For further confirmation, we have done Rietveld analysis of our XRD data which indicates system retains its original triclinic-\textit{I$\bar{1}$} structure down to 20 K. Figs. 6a, 6b and 6c show temperature dependent structural parameters $a$, $b$ and $c$ for present Sr$_2$FeIrO$_6$ as obtained from Rietveld analysis of XRD data. The evolution of lattice parameters with temperature is not though monotonic. With cooling, lattice parameters $a$, $b$ and $c$ decreases, however, Fig. 6 shows all the parameters show anomalous behavior across 45 K where $M(T)$ shows prominent peak. Moreover, lattice parameter $c$ also exhibits change in slope across 120 K. Nonetheless, though this material does not show structural phase transition across the magnetic phase transition, but we observe structural parameters show anomalous behavior across the magnetic phase transition.

\begin{figure}
\centering
		\includegraphics[width=8cm]{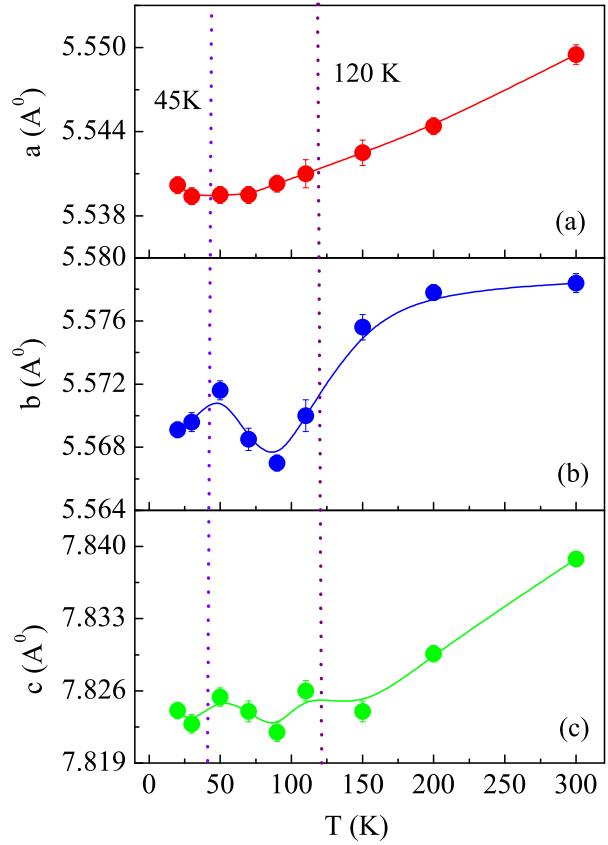}
\caption{(color online) Temperature variation in lattice parameters (a) $a$, (b) $b$ and (c) $c$ has been shown for Sr$_2$FeIrO$_6$. Lines are guide to eye.}
	\label{fig:Fig6}
\end{figure}

\subsection{Evolution of magnetic behavior in (Sr$_{1-x}$Y$_x$)$_2$FeIrO$_6$}

The effect of Y substitution on magnetization in present series is shown in Fig. 7. For low concentration of Y till $x$ = 0.1, not only the transition temperature $T_N$ shifts to higher temperature but the amount of bifurcation between $M_{FC}$ and $M_{ZFC}$ also increases. For $x$ $>$ 0.1, $T_N$ decreases and then again increases for $x$ = 0.2. Variation of $T_N$ with $x$ is shown in Fig. 9a. Fig. 8 shows temperature dependent inverse susceptibility data for (Sr$_{1-x}$Y$_x$)$_2$FeIrO$_6$ series. The $\chi^{-1}(T)$ for $x$ = 0.0, 0.05 and 0.1 in Fig. 8a and for $x$ = 0.15 and 0.2 in Fig. 8b show linear behavior in high temperature PM regime. The $\chi^{-1}(T)$ data are fitted with modified Curie-Weiss law (Eq. 1) which shows all the samples obey the CW behavior in PM state. As for parent $x$ = 0.0 material, we have also estimated $\theta_P$ and $\mu_{eff}$ for all the samples which are shown in Figs. 9b and 9c, respectively. The variation of $\theta_P$ with composition is similar to $T_N$ where it initially increases till $x$ = 0.1, and then decreases and increases. While $\theta_P$ remains negative for all the compositions but the nature of its change with $x$ suggest change of magnetic exchange interaction which may be an outcome of changing ionic state of Fe/Ir due to Y substitution. In particular, increase of $\theta_P$ from -169 K for $x$ = 0.0 to -15 K for $x$ = 0.1 suggests enhancement of ferromagnetic type exchange interaction by Y-substitution. The variation of $\mu_{eff}$ with $x$ is, however, opposite to $T_N$ and $\theta_P$. As discussed before, from $\mu_{eff}$ we could infer that Fe and Ir transition metals are in Fe$^{3+}$ and Ir$^{5+}$ state in parent material. The Y$^{3+}$ substitution for Sr$^{2+}$ will convert either Fe$^{3+}$ to Fe$^{2+}$ or Ir$^{5+}$ to Ir$^{4+}$. While in former case Fe$^{2+}$ with its 3$d^4$ ($S$ = 2) state will decrease overall $\mu_{eff}$ compared to Fe$^{3+}$ ($S$ = 5/2), and in later case converted Ir$^{4+}$ with 5$d^5$ ($J_{eff}$ = 1/2) state will increase effective $\mu_{eff}$ with respect to Ir$^{5+}$ ($J_{eff}$ = 0). Therefore, initial decrease of $\mu_{eff}$ till $x$ = 0.1 in Fig. 9c implies Y$^{3+}$ substitution converts Fe$^{3+}$ to Fe$^{2+}$ at lower concentration and then with increasing $x$, conversion of Ir$^{5+}$ to Ir$^{4+}$ takes place. 

\begin{figure}
\centering
		\includegraphics[width=8cm]{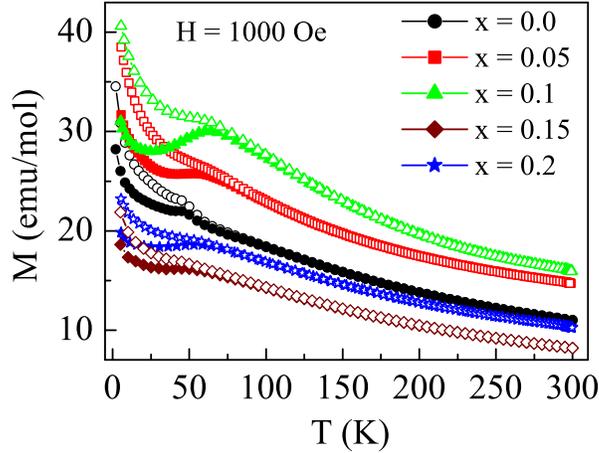}
\caption{(color online) DC magnetization data measured in 1000 Oe applied field following ZFC and FC protocol are shown as a function of temperature for (Sr$_{1-x}$Y$_x$)$_2$FeIrO$_6$ series. Filled and open symbols represent ZFC and FC magnetization data, respectively.} 
	\label{fig:Fig7}          
\end{figure}

\begin{figure}
\centering
		\includegraphics[width=8cm]{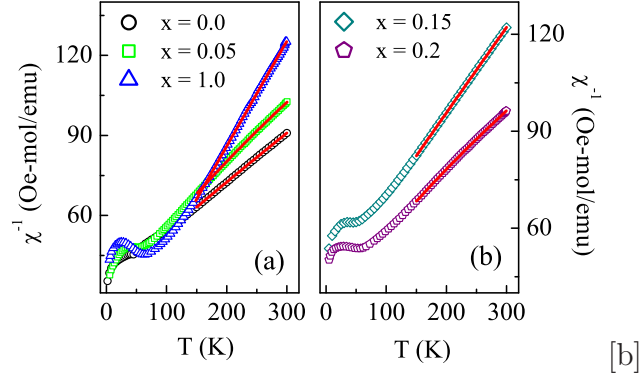}[b]
\caption{(color online) Inverse magnetic susceptibility ($\chi^{-1}$ = $(M/H)^{-1}$) deduced from ZFC magnetization data as shown in Fig. 7 are plotted for (Sr$_{1-x}$Y$_x$)$_2$FeIrO$_6$ series with $x$ = 0.0, 0.05 and 0.1 in (a) and $x$ = 0.15 and 0.2 in (b). Solid lines are due to fitting with modified Curie-Weiss behavior using Eq. 1.} 
	\label{fig:Fig8}
\end{figure}

\subsection{Magnetic field dependent magnetization study in (Sr$_{1-x}$Y$_x$)$_2$FeIrO$_6$}

Magnetic field ($H$) dependent magnetization data collected at 5 K up to field 70 kOe for (Sr$_{1-x}$Y$_x$)$_2$FeIrO$_6$ series have been shown in Fig. 10. For Sr$_2$FeIrO$_6$, $M(H)$ plot is almost linear till 45 kOe, then there is slight deviation from linearity where this linear $M(H)$ is a typical feature of AFM materials. There is no, however, sign of saturation till highest measuring field and we observe moment at 70 kOe is about $\mu_H$ = 0.256 $\mu_B$/f.u which is significantly lower than the calculated moment 5.0 $\mu_B$ per Fe$^{3+}$ ion. The $M(H)$ data shows small hysteresis with coercive field $H_c$ $\sim$ 590 Oe and remnant magnetization $M_r$ $\sim$ 2.8 $\times$ 10$^{-3}$ $\mu_B$/f.u. To understand the nature of magnetism, we have plotted the $M(H)$ data in form of $M^2$ vs $H/M$ which is commonly known as Arrott plot (see inset of Fig. 10).\cite{arrott} The significance of Arrott plot is that positive intercept due to straight line fitted in high field regime implies spontaneous magnetization or ferromagnetic type spin exchange interaction. On contrary, negative intercept implies an imaginary spontaneous magnetization which excludes the possibility of ferromagnetic ordering. Inset in Fig. 10 shows Arrott plot for parent Sr$_2$FeIrO$_6$ material where the negative intercept is suggestive of non-ferromagnetic nature of magnetic state which is in conformity with other magnetic measurements.

\begin{figure}
\centering
		\includegraphics[width=8cm]{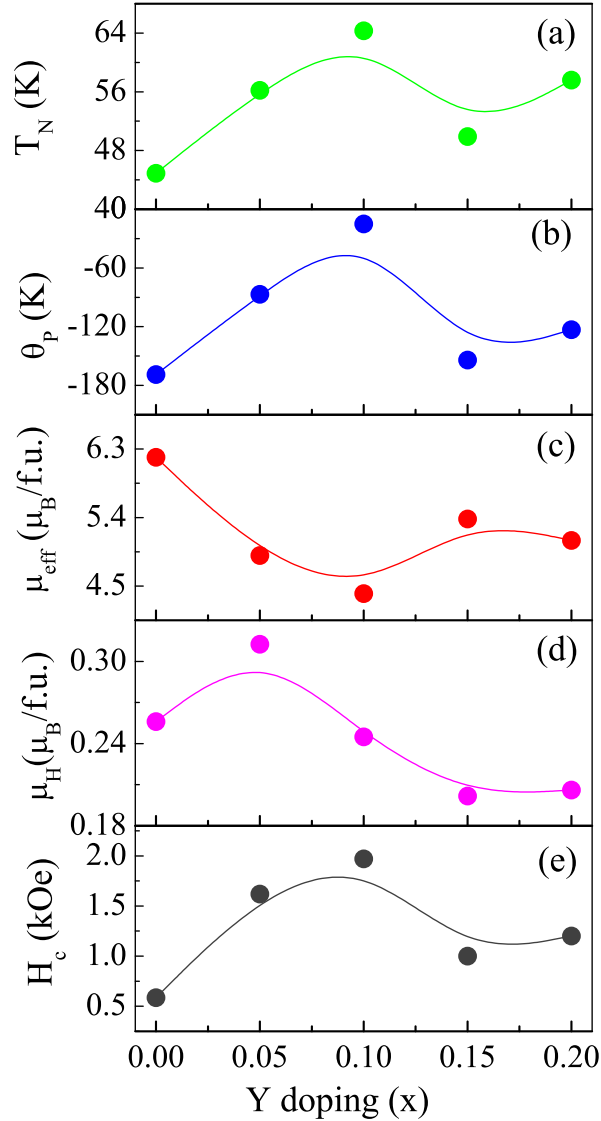}
\caption{(color online) Variation of (a) T$_N$ (b) $\theta_P$ (c) $\mu_{eff}$ (d) $\mu_H$ and (e) $H_c$ has been shown as a function of composition $x$ for the series (Sr$_{1-x}$Y$_x$)$_2$FeIrO$_6$.}
	\label{fig:Fig9}
\end{figure}

With progressive substitution of Y$^{3+}$, Fig. 10 shows there are changes in opening of $M(H)$ plot and moment value, $\mu_H$. For instance, we observe $H_c$ and $M_r$ increases to 1960 Oe and 8.7 $\times$ 10$^{-3}$ $\mu_B$/f.u. for $x$ = 0.1 material, respectively. The composition dependent $\mu_H$ and $H_c$ are shown in Figs. 9d and 9e, respectively. This initial increase of $H_c$ and $M_r$ again implies increase of ferromagnetism in the system which is in agreement with increase of $\theta_P$ as discussed earlier (Fig. 9b). This is also supported by the fact that negative intercept in Arrott plot decreases from -0.5 to -0.2 ($\mu_B$/f.u)$^2$ after Y substitution from $x$ = 0.0 to 0.1 value. 

\begin{figure}
\centering
		\includegraphics[width=8cm]{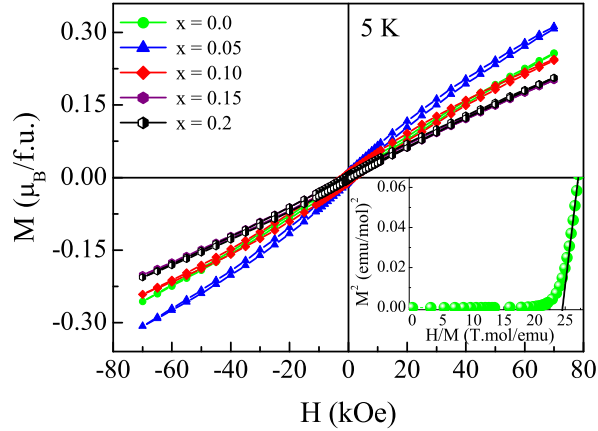}
\caption{(color online) Magnetic field dependent magnetization at 5 K are shown for (Sr$_{1-x}$Y$_x$)$_2$FeIrO$_6$ series. Inset shows Arrott plot ($M^2$ vs $H/M$) of magnetization data $M(H)$ for Sr$_2$FeIrO$_6$ at 5 K.}
	\label{fig:Fig10}
\end{figure}

\subsection{Evolution of electronic transport in (Sr$_{1-x}$Y$_x$)$_2$FeIrO$_6$}

Fig. 11 Shows temperature dependent electrical resistivity ($\rho$) for (Sr$_{1-x}$Y$_x$)$_2$FeIrO$_6$ series. Figure shows resistivity for $x$ = 0.0 parent material is highly insulating where resistivity increases drastically by about five orders cooling from 300 to 20 K. We also find there is a weak curvature change in $\rho(T)$ around 45 K, the temperature across which magnetic phase transition has been observed in Sr$_2$FeIrO$_6$ (Fig. 4a). Inset of Fig. 11 shows temperature derivative of resistivity, d$\rho$/d$T$ which prominently shows change is slope across 45 K. Substitution of Y shows non-monotonic changes in resistivity while all the samples remain to be insulating. The $\rho(T)$ data in Fig. 11 show resistivity increases substantially up to $x$ = 0.1 then it decreases for $x$ = 0.15 and 0.2. Nonetheless, the nature of changes of resistivity with Y concentration where we do observe a reverse in trend above $x$ = 0.1, is in conformity with structure and magnetization data.

We observe that nature of electron conduction in Sr$_2$FeIrO$_6$ follows thermally activated Mott's 3-dimensional (3D) variable range hopping model (VRH):\cite{mott}

\begin{eqnarray}
\rho = \rho_o \exp\left[\left(\frac{T_o}{T}\right)^4\right]	
\end{eqnarray}

Where $T_o$ is the characteristic temperature and can be expressed as:

\begin{eqnarray}
	T_o =\frac{18}{k_B N(E_F) \xi^3}
\end{eqnarray}

where $k_B$ is the Boltzmann constant, $N(E_F)$ is the density of states (DOS) at Fermi surface and $\xi$ is the localization length. Fig. 12a shows straight line fitting of the $\rho(T)$ data following Eq. 2 for $x$ = 0.0 material. The $\rho(T)$ data can be fitted in two different temperature regimes i.e., from lowest temperature to $\sim$ 45 K and then from 52 K to highest measuring temperature. This clearly shows though the nature of conduction mechanism remains same but the magnetic ordering around 45 K has influence on electronic conduction. Fig. 12a shows two linear regimes give two different slopes or two different values of $T_0$. The similar modification of $T_0$ as influenced by magnetic ordering has also been observed in layered iridate Sr$_2$IrO$_4$.\cite{bhatti} We obtain $T_0$ = 4.93 $\times$ 10$^5$ K and 0.16 $\times$ 10$^5$ K in low and high temperature regime, respectively. While these values of $T_0$ match reasonably with those for insulating materials but its change with temperature is quite intriguing. As seen in Eq. 3, change in $T_0$ could be realized through either change in DOS ($N(E_F)$) or change in $\xi$. While the modification in $N(E_F)$ with onset of magnetic ordering is quite unlikely, we believe change in $\xi$ may cause modification of $T_0$. It can be recalled that we observe a sudden modification in structural parameters across magnetic phase transition around 45 K (Fig. 6). This coupled with the fact that $\rho(T)$ also shows (weak) change in curvature around 45 K, the change of $T_0$ may be linked to change in localization length $\xi_0$.   

\begin{figure}
\centering
		\includegraphics[width=8cm]{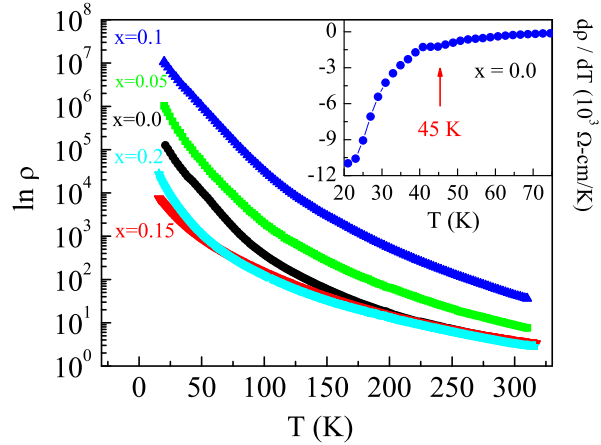}
\caption{(color online) Temperature variation in resistivity has been shown for (Sr$_{1-x}$Y$_x$)$_2$FeIrO$_6$ series in semilogarithmic plot. Inset shows temperature derivative of resistivity (d$\rho$/dT) for Sr$_2$FeIrO$_6$ as a function of temperature.}
	\label{fig:Fig11}
\end{figure}

For the Y substituted samples, however, we observe that $\rho(T)$ data can not be fitted with VRH model. Instead, we find temperature dependent resistivity data, for all the doped samples, follow power law behavior, 

\begin{eqnarray}
	\rho = \rho_o T^{-n}
\end{eqnarray}

Fig. 12b shows $\rho(T)$ data in form of $\log \rho$ vs $\log T$ where the straight lines are due to fitting of data using Eq. 4. We find that power law behavior (Eq. 4) is fairly obeyed in higher temperature regime above the magnetic phase transition temperature $T_N$. However, at low temperature below $T_N$ data in Fig. 12b is not very linear. In Table I, we have given the exponent $n$ for all the samples.

\subsection{Discussions}

From above discussions, it is clear that Sr$_2$FeIrO$_6$ has prominent magnetic phase transition around 45 K as indicated by a kink and sizable bifurcation between M$_{ZFC}$ and M$_{FC}$. An additional though weak signature of magnetic phase change has been observed around 120 K (Fig. 4). While understanding the detail nature of these magnetic state or interaction requires microscopic experimental investigations, it can be pointed out that similar kind of different AFM states through change in spin structure is also evident at low temperature in other 3$d$-5$d$ type DP systems such as, Sr$_2$FeOsO$_6$ and Sr$_2$CoOsO$_6$.\cite{avijit, williams,yan,avijit1,sudipta,morrow} Substitution of Y in present (Sr$_{1-x}$Y$_x$)$_2$FeIrO$_6$ series has an interesting effect where the structural parameters, magnetic parameters and electronic state evolves with $x$, however, the nature of change reverses for $x$ $>$ 0.1. Analysis of our magnetization data, which is also supported by previous MS, XAS and XMCD data, imply that in parent Sr$_2$FeIrO$_6$ the transition metals are in state of Fe$^{3+}$ (3$d^5$) and Ir$^{5+}$ (5$d^4$) those are magnetic ($S$ = 5/2) and nonmagnetic ($S$ = 0), respectively. Therefore, in an ideal disorder-free situation (Fig. 2) there are basically two channels of magnetic interaction; one is between Fe1 and Fe2 which is linear along $c$-axis and mediated through Ir i.e., Fe1-O-Ir-O-Fe2 channel and another one is between Fe1 and Fe3 which follows a nonlinear path of Fe1-O-O-Fe3. In former case the separation between Fe ions is higher ($\sim$ 7.84 \AA) than in later situation ($\sim$ 5.55 \AA). While strength of both these interactions can not be understood with present data set but these magnetically active Fe ions would engage in superexchange interaction establishing long-range AFM state. Moreover, it remains interesting to understand whether these two channels of magnetic interaction are related to two transition temperatures around 45 and 120 K.

\begin{figure}
\centering
		\includegraphics[width=8cm]{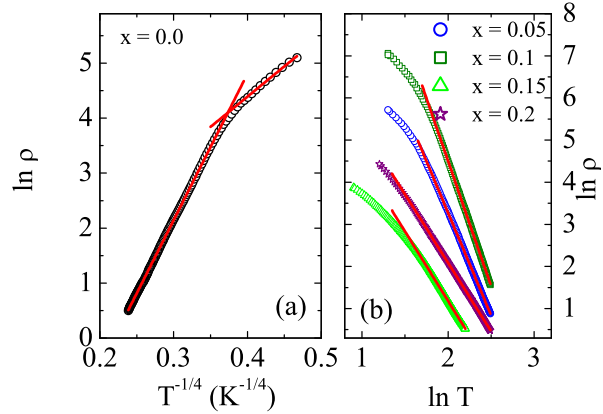}
\caption{(color online) (a) Resistivity data have been plotted in form of $\ln \rho$ vs $T^{-1/4}$ for Sr$_2$FeIrO$_6$. Straight lines are due to fitting of data following Eq. 2. (b) Resistivity data again plotted in form of $\ln \rho$ vs $\ln T$ for (Sr$_{1-x}$Y$_x$)$_2$FeIrO$_6$ series with $x$ $>$ 0. Straight lines are due to fitting of data following Eq. 4.}
	\label{fig:Fig12}
\end{figure}
        
With substitution of Y, initial conversion of Fe$^{3+}$ ($S$ = 5/2) in to Fe$^{2+}$ ($S$ = 2) will have influence on the magnetic state. Although, Fe ions would still engage in AFM-type superexchange interaction but the mismatch in its moment will result in rather ferrimagnetic type interaction which is also supported by the fact that low temperature moment $\mu_H$ increases, $\theta_P$ becomes less negative and $\mu_{eff}$ decreases (Fig. 9). The prominent example of ferrimagnetic exchange interaction between Fe$^{3+}$ and Fe$^{2+}$ is Fe$_3$O$_4$. Above $x$ = 0.1, we believe conversion of Ir$^{5+}$ to Ir$^{4+}$ dominates where the later being magnetic would participate in AFM superexchange interaction with both neighboring Ir$^{4+}$ as well as with Fe ions. This would promote AFM interaction as seen by reversing behavior of related parameters in Fig. 9.  

\begin{table}
\caption{\label{tab:table 1} Exponent $n$ obtained from fitting resistivity data with Eq. 4  are shown with doping concentration ($x$) for (Sr$_{1-x}$Y$_x$)$_2$FeIrO$_6$ series.}
\begin{indented}
\item[]\begin{tabular}{cc}
\hline
Sample ($x$)  & exponent ($n$)\\
\hline
\hline
0.05 &   4.8\\
0.10 &   5.9\\
0.15 &   3.3\\
0.20 &   3.25\\ 
\hline
\end{tabular}
\end{indented}
\end{table} 

\section{Conclusion}

In conclusion, we have prepared polycrystalline double perovskite samples of series (Sr$_{1-x}$Y$_x$)$_2$FeIrO$_6$ with $x$ = 0, 0.05, 0.1, 0.15 and 0.2. Analysis of XRD data shows all the samples are in single phase and crystallize in triclinic-\textit{I$\bar{1}$} crystal symmetry. The magnetization data for parent Sr$_2$FeIrO$_6$ show a AFM type magnetic transition around 45 K, however, a closer view  of data reveals onset of (weak) magnetic transition around 120 K. Given that similar two transition temperatures (T$_N$) related to change in spin structure of AFM state has been observed in other 3$d$-5$d$ systems, the present Sr$_2$FeIrO$_6$ material requires detailed investigation using microscopic tool. Temperature dependent structural investigation shows no change of structural symmetry down to low temperature, although lattice parameters show unusual changes around magnetic transitions. The parent Sr$_2$FeIrO$_6$ is an insulator throughout the temperature range, and with Y substitution electronic state remains insulator though resistivity changes. The nature of electronic charge conduction has been found to follow 3D variable range hopping model and power law behavior for parent Sr$_2$FeIrO$_6$ and doped materials, respectively. Substitution of Y in (Sr$_{1-x}$Y$_x$)$_2$FeIrO$_6$ has an overall interesting effect where the evolution of structural, magnetic and electronic properties show reverse trend for $x$ $>$ 0.1. We hope our results will inspire more similar investigations of theoretical calculations as well as experimental studies employing other doping elements.    

\ack{We are thankful to AIRF, JNU for low temperature x-ray diffraction measurement and Manoj Pratap Singh for the help in the measurement. We acknowledge IIT Delhi for magnetization measurements. KCK acknowledges University Grant Commission, India for SRF fellowship.}

\end{document}